\documentclass[letter,longauth]{aa}
\usepackage{txfonts}
\usepackage{graphicx}
\usepackage{color}
\usepackage{natbib}
\begin{document}

\title{A HIFI preview of warm molecular gas 
around $\chi$ Cyg : first detection of H$_{2}$O emission
toward an S-type AGB star
\thanks{Herschel is an ESA space observatory with science instruments 
provided by European-led Principal Investigator consortia and with 
important participation from NASA.}
}
\titlerunning{A HIFI preview of warm molecular gas around $\chi$ Cyg}
\author{K. Justtanont \inst{1}
\and L. Decin \inst{2,3}
\and F.~L. Sch\"oier \inst{1}
\and M. Maercker \inst{4,22}
\and H. Olofsson \inst{1,5}
\and V. Bujarrabal \inst{6}
\and A.P. Marston \inst{7}
\and D. Teyssier \inst{7}
\and J. Alcolea \inst{8}
\and J. Cernicharo \inst{9}
\and C. Dominik \inst{3,10}
\and A. de Koter \inst{3,11}
\and G. Melnick \inst{12}
\and K. Menten \inst{13}
\and D. Neufeld \inst{14}
\and P. Planesas \inst{6,15}
\and M. Schmidt \inst{16}
\and R. Szczerba \inst{16}
\and R. Waters \inst{3,2}
\and Th. de Graauw \inst{17}
\and N. Whyborn\inst{17}
\and T. Finn\inst{18}                      
\and F. Helmich\inst{19}
\and O. Siebertz\inst{20}
\and F. Schm\"{u}lling\inst{20}
\and V. Ossenkopf\inst{20}
\and R. Lai\inst{21}
}
\institute{Onsala Space Observatory, Chalmers University of Technology, 
Dept. Radio \& Spece Science, SE--439 92 Onsala, Sweden
\email{kay.justtanont@chalmers.se}
\and Instituut voor Sterrenkunde, Katholieke Universiteit Leuven, 
Celestijnenlaan 200D, 3001 Leuven, Belgium
\and Sterrenkundig Instituut Anton Pannekoek, University of Amsterdam,
Science Park 904, NL-1098 Amsterdam, The Netherlands
\and University of Bonn, Argelander-Institut f\"{u}r Astronomie, 
Auf dem H\"{u}gel 71, D-53121 Bonn, Germany
\and Department of Astronomy, AlbaNova University Center, Stockholm
University, SE--10691 Stockholm, Sweden
\and Observatorio Astron\'omico Nacional. Ap 112, E-28803
Alcal\'a de Henares, Spain
\and European Space Astronomy Centre, ESA, P.O. Box 78, E-28691
Villanueva de la Ca\~nada, Madrid, Spain
\and Observatorio Astron\'omico Nacional (IGN), Alfonso XII N$^{\circ}$3,
E-28014 Madrid, Spain
\and CAB, INTA-CSIC, Ctra de Torrej\'on a Ajalvir, km 4,
28850 Torrej\'on de Ardoz, Madrid, Spain
\and Department of Astrophysics/IMAPP, Radboud University Nijmegen,   
Nijmegen, The Netherlands
\and
Astronomical Institute, Utrecht University,
Princetonplein 5, 3584 CC Utrecht, The Netherlands
\and Harvard-Smithsonian Center for Astrophysics, Cambridge, MA 02138, USA
\and Max-Planck-Institut f{\"u}r Radioastronomie, Auf dem H{\"u}gel 69,
D-53121 Bonn, Germany
\and Johns Hopkins University, Baltimore, MD 21218, USA
\and Joint ALMA Observatory, El Golf 40, Las Condes, Santiago, Chile
\and N. Copernicus Astronomical Center, Rabia{\'n}ska 8, 87-100 Toru{\'n}, 
Poland
\and Atacama Large Millimeter/Submillimeter Array, Joint 
ALMA Office, Santiago, Chile
\and Experimental Physics Dept., National University of Ireland 
Maynooth, Co. Kildare. Ireland
\and SRON Netherlands Institute for Space Research, Landleven 
12, 9747 AD Groningen, The Netherlands
\and KOSMA, I. Physik. Institut, Universit\"{a}t zu K\"{o}ln, 
Zülpicher Str. 77, D 50937 K\"{o}ln, Germany
\and Northrop Grumman Aerospace Systems, 1 Space Park, Redondo
Beach, CA 90278 U.S.A
\and European Southern Observatory, Karl Schwarzschild Str. 2, Garching bei
M\"{u}nchen, Germany
}
\date{Received 31 May 2010 / Accepted 9 June 2010 }
\abstract{}
{A set of new, sensitive, and spectrally resolved, sub-millimeter
line observations are used to probe the warm circumstellar gas around
the S-type AGB star $\chi$ Cyg. The observed lines involve high
rotational quantum numbers, which, combined with previously obtained
lower-frequency data, make it possible to study in detail the chemical and 
physical properties of, essentially, the entire
circumstellar envelope of $\chi$ Cyg.}
{The data were obtained using the HIFI instrument aboard Herschel,
whose high spectral resolution provides valuable information about
the line profiles. Detailed, non-LTE, radiative transfer modelling,
including dust radiative transfer coupled with a dynamical model, has
been performed to derive the temperature, density, and velocity
structure of the circumstellar envelope.}
{We report the first detection of circumstellar H$_{2}$O rotational 
emission lines
in an S-star. Using the high-$J$ CO lines to derive the parameters
for the circumstellar envelope, we modelled both the 
ortho- and para-H$_{2}$O lines. Our modelling results are
consistent with the velocity structure expected for a dust-driven wind.
The derived total H$_{2}$O abundance 
(relative to H$_{2}$) is $(1.1 \pm 0.2)\times10^{-5}$, much lower than that
in O-rich stars.
The derived ortho-to-para ratio of 2.1$\pm$0.6 is close to the
high-temperature equilibrium limit, consistent with H$_{2}$O being formed
in the photosphere.
}
{}
\maketitle

\section{Introduction}

Observations of the 
dust and gas components in the
circumstellar envelopes (CSEs) around asymptotic 
giant branch (AGB) stars have been carried out at different wavelengths.
Observations in the infrared trace the dust 
as well as the warm molecular layer close to the stellar photosphere
(e.g., \citealt{just96,aoki99,schoier02}).
Submillimeter and radio observations of trace molecules have been used to
study the cooler outer parts of the CSEs (e.g., \citealt{knapp85,
schoier00, kemper03}).
To bridge this gap, the Infrared Space Observatory (ISO) was used to
observe a large number of AGB stars up to almost 200~${\mu}$m, 
and in O-rich 
stars, a number of H$_{2}$O emission lines were detected
\citep{barlow96, neufeld96}.
However, the circumstellar far-infrared
lines were unresolved. Hence, crucial information
about the line profiles remained unknown. 

Water is an important molecule in CSEs as it is thought to be one
of the main cooling agents in the wind. It is also expected to be a
good probe of the inner regions of the CSE where the gas is accelerated.
However, to fully explore the
potential of H$_{2}$O lines as a probe of the circumstellar gas,
a full radiative transfer has to be performed. Owing to difficulties
in calculating accurate collisional rates coupled with the very high optical
depth of the H$_{2}$O lines in the inner region of the CSE, 
slow progress has been made. Nevertheless, calculations 
of the heating and cooling in the CSEs of O-rich stars suggest 
that H$_{2}$O dominates the
cooling in most parts of the envelope until it is photodissociated
by interstellar UV photons \citep{gold76,
just94, maercker08, maercker09, decin10}, and that some lines should
come mainly from the acceleration zone.
Eventually, spectrally resolved circumstellar H$_{2}$O lines
were observed by two space missions dedicated to search for cosmic
water line emission: SWAS and Odin. Both missions were
able to detect the ground-state line of H$_{2}$O at 557 GHz in a number
of AGB stars \citep{harwit02, melnick01, just05, hase06, maercker09}. 
It was shown that not only the line intensity, but
also the line profile is crucial for interpreting the data correctly.
 
In 2009, the ESA-Herschel Space Observatory \citep{pilbratt10} 
was launched with the
Heterodyne Instrument for the Far-Infrared (HIFI\footnote{
HIFI has been designed and built by a consortium of institutes and 
university departments from across Europe, Canada and the United States 
under the leadership of SRON Netherlands Institute for Space
Research, Groningen, The Netherlands and with major contributions 
from Germany, France and the US. Consortium members are: Canada: CSA, 
U.Waterloo; France: CESR, LAB, LERMA, IRAM; Germany:
KOSMA, MPIfR, MPS; Ireland, NUI Maynooth; Italy: ASI, IFSI-INAF, 
Osservatorio Astrofisico di Arcetri-
INAF; Netherlands: SRON, TUD; Poland: CAMK, CBK; Spain: Observatorio 
Astron\'{o}mico Nacional (IGN),
Centro de Astrobiolog\'{i}a (CSIC-INTA). Sweden: 
Chalmers University of Technology - MC2, RSS \& GARD;
Onsala Space Observatory; Swedish National Space Board, 
Stockholm University - Stockholm Observatory;
Switzerland: ETH Zurich, FHNW; USA: Caltech, JPL, NHSC.
}, \citealt{degraauw10}),
which aims to study H$_{2}$O line emission in different environments
in our Galaxy and beyond. HIFI offers the opportunity to
study the warm molecular layers in CSEs of AGB stars in great detail, e.g.,
the high spectral resolution and wide spectral coverage allow 
a detailed study of the gas dynamics. 

As part of the guaranteed time programme HIFISTARS (P.I.: V. Bujarrabal),
the S-star (C/O\,$\approx$\,1)
$\chi$ Cyg was selected for study. Distance estimates
range from 150 pc \citep{knapp03} to 180 pc \citep{leeuwen07}.
The star exhibits SiO masers (e.g., \citealt{olofsson81,
schwartz82, alcolea92}), but no H$_{2}$O maser emission has been found
\citep{menten95, shintani08}. 
Being nearby and bright, $\chi$ Cyg has been observed using 
interferometric techniques in both the optical
\citep{lacour09} and the infrared \citep{tevousjan04}.
In this Letter, 
we briefly describe the observations in Sect.~2, we discuss the 
modelling of the observed molecular emission in Sect.~3, and
present our results and conclusions in Sect.~4.

\section{Observations}

\begin{table}
\caption{HIFI observations of $\chi$ Cyg.}
\label{tab-obs}
\begin{tabular}{llrrr}
\hline \hline
  Molecule  &  Transition & $\nu$ (GHz) & $E_{\rm up}$ (K)& $I^{a}$
( K km s$^{-1}$) \\
\hline 
CO        &  $J$=6--5           &  691.473 & 116 & 15.3   \\
CO        &  $J$=10--9          & 1151.985 & 304 & 13.6 \\
CO        &  $J$=16--15         & 1841.346 & 752 & 13.5 \\
o-H$_{2}$O & 1$_{10} - 1_{01}$ & 556.936  &  61 & 3.0   \\
p-H$_{2}$O & 2$_{11} - 2_{02}$ & 752.033  & 137 &  4.2   \\
p-H$_{2}$O & 2$_{02} - 1_{11}$ & 987.927  & 101 & 7.3   \\
p-H$_{2}$O & 1$_{11} - 0_{00}$ & 1113.343 &  53 & 8.0   \\
o-H$_{2}$O & 3$_{12} - 2_{21}$ & 1153.127 & 249 & 7.5   \\
o-H$_{2}$O & 3$_{21} - 3_{12}$ & 1162.912 & 305 & 1.5   \\
o-H$_{2}$O & 3$_{03} - 2_{12}$ & 1716.770 & 197 & 18.1   \\
\hline
\end{tabular}
Note to the table: $^{a}$The absolute calibration accuracy 
is between 10\% to 30\% (see text).
\end{table}

The HIFI data were obtained using the dual-beam-switching
mode \citep{roelfsema10}
with a throw of 3$^{\prime}$ and slow (0.5-1Hz) chopping
in March-April 2010. A total of 8 frequency
settings with a total of 10 lines detected are being reported in this paper. 
The targeted lines were selected to cover
a wide range of excitation temperature, exploring different
regions of the CSE. 
As backend, the wide-band spectrometer (WBS)
covering a region of 4 GHz with a resolution of 1.1 MHz was used.
More details about these observations can be found in \citet{bujar10}.
The data were calibrated using the standard pipeline for
Herschel, HIPE 
version 2.8. Only the H-polarization data are presented
here because the V-polarization data are noisier especially for the
high frequency lines.
We subtracted the baseline using a first or second order polynomial, 
except for the H$_{2}$O transition at 1716 GHz, where this line 
is affected by standing waves
and, consequently, the baseline was fitted using a high order polynomial
(see Sect. 3.2).

We detected high rotational transitions of CO as well
as the first detection of rotational H$_{2}$O in an S-star.
All of the observed lines listed in
Table~\ref{tab-obs} were detected. The frequency $\nu$ 
in GHz and the energy
of the upper level $E_{\rm up}$ in K are given along with the 
integrated line intensity, $I=\int\!\!T_{\rm mb}\rm d\upsilon$ in K\,km\,s$^{-1}$.
The spectra were corrected for the main beam efficiency, i.e.,
$\eta_{\rm mb} = 0.72~\rm exp(-(\nu(\rm GHz)/6000)^2)$.
The absolute calibration accuracy ranges from 10\% for the lowest
frequency line  to 30\% for the high frequency ($>$ 1 THz) lines.

\section{Modelling HIFI lines}

We started the analysis by fitting the spectral energy distribution of the 
CSE (assumed to be spherically symmetric) using Dusty \citep{ivezic97} to
derive the dust mass-loss rate.
Based on this, we solved the gas-dust drag equation,
assuming that both are momentum coupled, to derive the velocity 
structure using the observed terminal gas velocity as a constraint, 
as shown in Fig.~\ref{fig-temp}. This also provides a so-called
dynamical mass-loss rate estimate, 4.9$\times 10^{-7}$ M$_{\odot}$ yr$^{-1}$ in
the case of $\chi$\,Cyg (see e.g.,
\citealt{ramstedt08}).
The gas velocity law is fed into the CO radiative transfer model,
where the line fluxes and shapes
are computed and compared to the observations.
At the same time, the heating by dust grains and the  cooling by 
CO lines are calculated and the resulting temperature structure
obtained (e.g., \citealt{just94, crosas97, decin06, vdtak07, ramstedt08}).
This provides a circumstellar model, including a mass-loss-rate estimate
based on the CO line modelling, which is used to model the H$_{2}$O line
emission and its contribution to the cooling.
The H$_{2}$O cooling is then used to recalculate the gas kinetic temperature
in the CO model,
and the process is iterated until good fits to the observed CO
and H$_{2}$O lines
are obtained. 

\subsection{Modelling of the CO lines}

\begin{figure}
\centering{
\includegraphics[width=6cm,angle=-90]{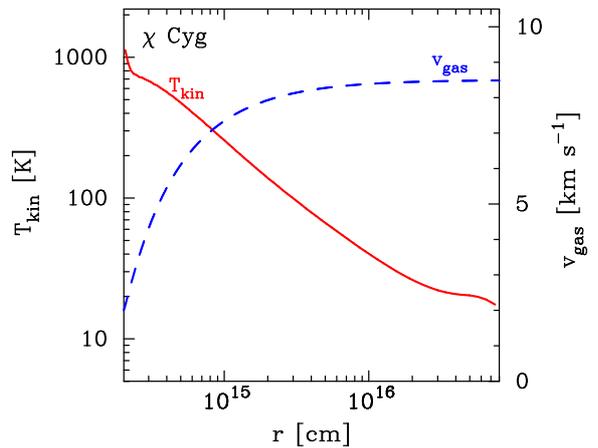}
\caption{The derived gas temperature structure (solid line)
and the gas expansion velocity (dashed line) of the CSE of $\chi$ Cyg.
}
}
\label{fig-temp} 
\end{figure}

The Monte Carlo code developed by \citet{schoier01} was used to model
the observed CO lines. The
molecular data were taken from the Hitran database
\citep{rothman09} and the
collisional rate coefficients from \citet{yang10} for the 
41 lowest rotational levels in the $v=0$ vibrational state.
We fitted the line shapes
and fluxes for the low-$J$ lines obtained from ground-based observations,
as well as the interferometric data for the $J$\,=\,1--0 and 2--1 lines
to more tightly constrain the size of the CO envelope (Sch\"{o}ier et al, 
in preparation). 
The parameters are listed in Table~\ref{tab-model}. The 
gas kinetic 
temperature distribution is shown in Fig.~\ref{fig-temp}.

\begin{figure*}
\sidecaption
\includegraphics[width=12cm]{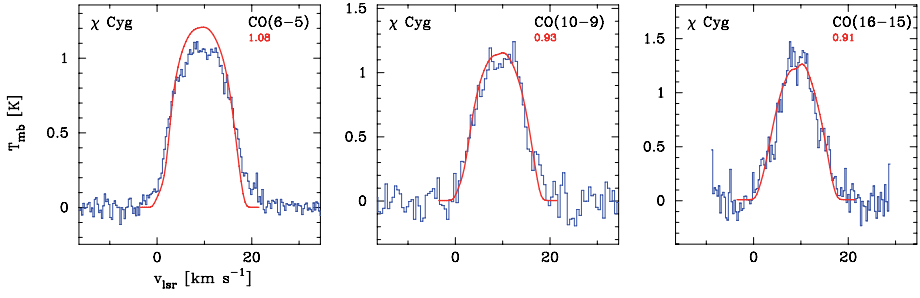}
\caption{The model fits (smooth lines) to
the HIFI CO $J$\,=\,6--5, 10--9 and 16--15 lines (histogram). 
The scaling factor (used to scale the model result to fit the 
observed integrated intensity) is given for each line,
showing the goodness of the fit.}
\label{COfits}
\end{figure*}

\begin{table}
\caption{Parameters used in the modelling of the line emission.}
\label{tab-model}
\begin{tabular}{ l l}
\hline 
Distance                         & 150 pc \\
Stellar effective temperature ($T_{\rm eff}$) & 2600 K \\
Stellar luminosity ( $L_{*}$)     & 7.5$\times 10^{3}$ L$_{\odot}$ \\
Gas terminal velocity ($v_{\rm exp}$)     & 8.5 km s$^{-1}$ \\
Inner radius of the shell ($R_{\rm in}$) & 2$\times 10^{14}$ cm\\
Gas mass-loss rate  ($\dot{M}$)    & 7$\times 10^{-7}$ M$_{\odot}$ yr$^{-1}$ \\
CO abundance (CO/H$_{2}$)          & 6$\times 10^{-4}$  \\
ortho-H$_{2}$O abundance (o-H$_{2}$O/H$_{2}$)  &  7.5$\times 10^{-6}$ \\
para-H$_{2}$O abundance (p-H$_{2}$O/H$_{2}$)  &  3.6$\times 10^{-6}$ \\
\hline
\end{tabular}
\end{table}

We assume a distance of 150 pc \citep{knapp03}, 
and the resulting mass-loss rate is 7$\times 10^{-7}$ M$_{\odot}$ yr$^{-1}$ 
using an adopted CO abundance of 6$\times 10^{-4}$ (relative to H$_{2}$, 
see Table~\ref{tab-model}), which provide the best fits to both
the line profile and line intensities.
The uncertainty in the mass-loss rate is of the order 50\%.
This mass-loss rate agrees well with the
dynamical mass-loss rate (certainly to within the errors in the 
input parameters).
A comparison of the model fits and the HIFI observations
can be seen in Fig~\ref{COfits}. The models have been scaled
to match the line intensities and the scaling factor (given in 
each panel of the figure) is a measure of the goodness of fit.
The high rotational line at $J$\,=\,16--15 
is noticeably narrower than the lower-level lines observed with
HIFI and ground-based instruments (e.g., \citealt{knapp98}),
indicating that this line originates in a region where the
gas is still being accelerated. 
The estimated outer CO radius is 4$\times 10^{16}$ cm.

\subsection{Modelling of the H$_{2}$O lines}

\begin{figure*}
\centering
\includegraphics[width=17cm]{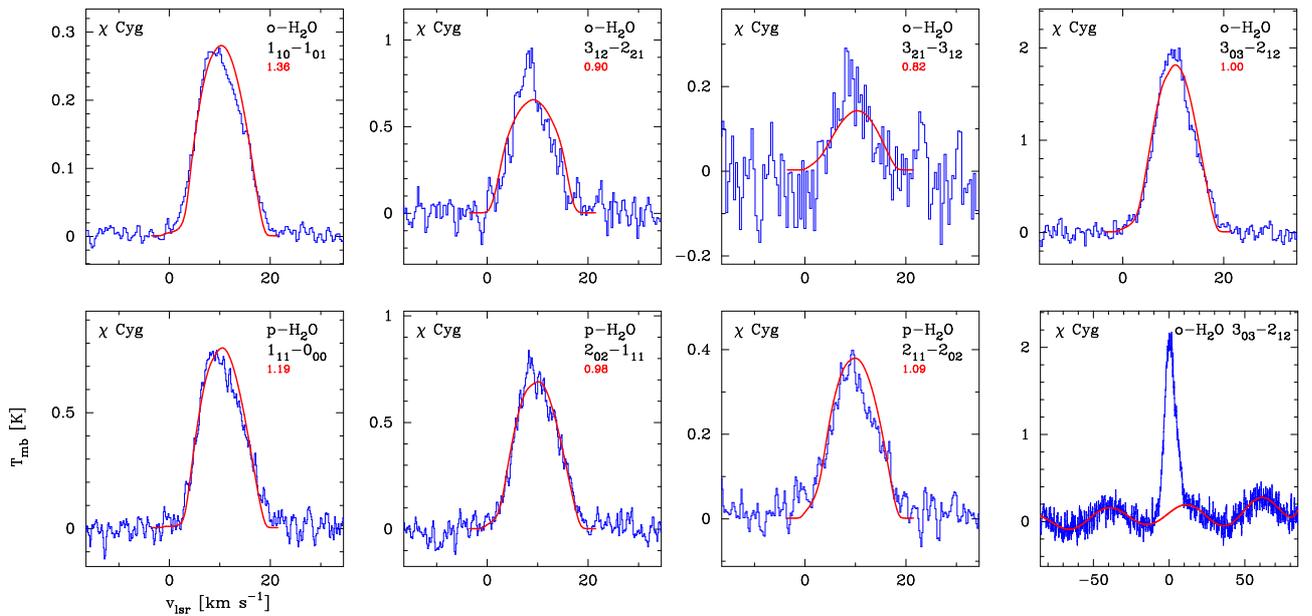}
\caption{ 
The model fits (smooth lines) to
the HIFI ortho- and para-H$_{2}$O lines (histogram). 
The scaling factor (used to scale the model result to fit the observed 
integrated intensity) is given for each line,
showing the goodness of the fit. 
The pipeline-reduced data of the 3$_{03} - 2_{12}$ 
line are plotted along with the fitted baseline at the bottom right.
This line is affected by standing waves within HIFI.
}
\label{H2Ofits}
\end{figure*}

To calculate the strength and shape of the circumstellar
H$_{2}$O lines,
we apply the parameters derived from our CO line modelling (mass-loss
rate, gas temperature and density structure, and gas velocity law)
to the radiative transfer model for
H$_{2}$O based on an accelerated lambda iteration (ALI) code
\citep{just05, maercker08, maercker09}. We used the molecular data
from \citet{rothman09} and the collisional cross-sections
from \citet{faure07} for the lowest 45 levels of ortho- and
para-H$_{2}$O. The radiative excitation due to the absorption in the 
$\nu_{2}$ bending and $\nu_{3}$ stretching modes is included.
The latter has been found to have a non-negligible effect
in the low mass-loss-rate case \citep{maercker09}.
The outer radius of the H$_{2}$O shell was derived  
using the model results of \citet{netzer87}, i.e., 3.6$\times 10^{15}$ cm.
Since the ortho- and para-species are expected
to be independent,
we model the two species separately, using the same circumstellar 
input values and estimate the two abundances independently. 
Using the temperature and velocity structure from the CO
modelling (Fig~\ref{fig-temp}), we calculate the best-fit model to the 
H$_{2}$O lines. All the lines are found to be sub-thermally excited. As 
mentioned above, both CO and H$_{2}$O line cooling is taken into account.

We present the fits to four lines of ortho-H$_{2}$O and three 
lines of para-H$_{2}$O observed with HIFI in Fig.~\ref{H2Ofits}. 
These lines span upper energies from 60 to 300\,K so the lines probe 
the cool part of the CSE as well as the warmer inner part. 
From Figs.~\ref{fig-temp} and \ref{cool}, it can be seen that
H$_{2}$O lines originate well within the acceleration zone.
All lines are of reasonably high signal-to-noise ratio, including the 
3$_{03} - 2_{12}$ line, 
the highest frequency line, where the spectrum is affected by
standing waves inside HIFI. For this line, we used a higher order Chebyshev
polynomial for the baseline subtraction (bottom right panel of 
Fig~\ref{H2Ofits}).

Our results are consistent with the velocity structure of a dust-driven 
wind as can be seen in good fits to both the CO and H$_{2}$O
line profiles (Figs.~\ref{COfits} and \ref{H2Ofits}, respectively). 
Unlike the case
for IK Tau \citep{decin10, decin10h}, no modification to the dynamical 
calculation is required.

\section{Results and conclusions}

The observed HIFI lines are reliable probes of the inner CSE as the 
high-energy lines probe the wind in the acceleration zone.
Both the velocity and density structures are tightly constrained using
the HIFI lines. The abundance of CO used is 6$\times 10^{-4}$, intermediate to
those usually adopted for O- and C-rich CSEs. The derived ortho- and 
para-H$_{2}$O 
abundances are significantly lower, (7.5$\pm1.4)\times 10^{-6}$ and 
(3.6$\pm0.5)\times 10^{-6}$, respectively (Table~\ref{tab-model}).
These values are well below the limits for O-rich AGB stars of $>$\,10$^{-4}$
\citep{just05, maercker08, maercker09} 
consistent with $\chi$ Cyg being an S-star of C/O 
very close to unity. 
From our modelling, assuming that all carbon is locked
up in CO (i.e., C/H = 3$\times 10^{-4}$) and the oxygen is locked up in both
CO and H$_{2}$O (i.e., O/H = 3$\times 10^{-4}$ + 5.5$\times 10^{-6}$), 
our derived C/O is $\leq$ 0.98, given that a small fractional abundance
of the oxygen is in dust grains. This value is slightly higher than
that of 0.95, assumed by \citet{duari00}.
A non-thermal equilibrium chemistry model 
for S-stars (C/O\,=\,0.98)
predicts an H$_{2}$O abundance of 10$^{-4}$ at the stellar photosphere,
falling off to a few 10$^{-6}$ at 5 R$_{*}$ \citep{cherchneff06},
an order of magnitude lower than our value.

In the thermal equilibrium (TE) limit at high temperature, the expected
ortho-to-para ratio is 3. 
Our derived 
ortho-to-para ratio is 2.1$\pm$0.6, close to the high-temperature TE value.
The reported ortho-to-para ratio in CSEs of O-rich
stars vary from 1 in W Hya with a large
uncertainty \citep{barlow96} to 3 in IK Tau
\citep{decin10h}. Our result is consistent with H$_{2}$O molecules 
being formed under thermal equilibrium conditions in
the warm and dense stellar photosphere.

\begin{figure}
\centering{
\includegraphics[width=6.5cm,angle=-90]{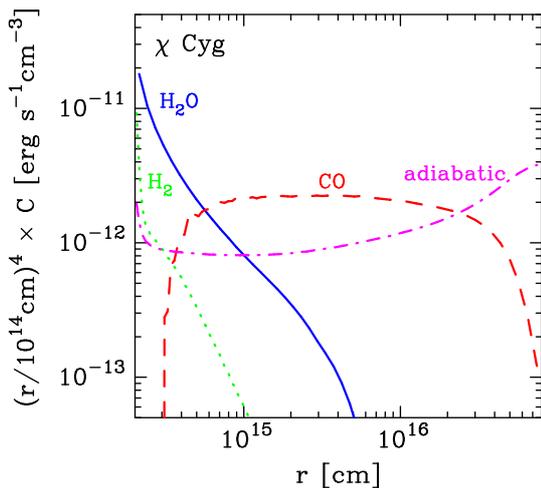}
\caption{The cooling rates due to different process : rotational cooling by
H$_{2}$O (solid), CO (dashed) lines, vibrational cooling by
H$_{2}$ (dotted) line, and adiabatic cooling (dot-dash).}
}
\label{cool}
\end{figure}

Given the low total H$_{2}$O abundance of $(1.1\pm0.2)\times 10^{-5}$,
it is clear from our analysis
that the dominating cooling agent in the
CSE of $\chi$ Cyg is CO (Fig.~\ref{cool}). 
Vibrationally excited H$_{2}$ and rotationally excited
H$_{2}$O contribute only in the innermost ($r$\,$<$\,10$^{15}$ cm) part 
of the CSE while CO line cooling 
extends further out until CO is photodissociated 
by the external UV field. Adiabatic cooling dominates only in 
the outermost part of the CSE.
The derived H$_{2}$O abundance, although much
lower than in O-rich stars, is higher than that observed in C-stars,
IRC+10216 \citep{melnick01} and V Cyg
\citep{neufeld10}, consistent with AGB stars of S-type being chemically 
intermediate between O-rich and C-rich AGB stars.

\begin{acknowledgements}
HCSS / HSpot / HIPE is a joint development (are joint developments) 
by the Herschel Science Ground Segment Consortium, consisting of ESA, 
the NASA Herschel Science Center, and the HIFI, PACS and SPIRE consortia.
K.J., F.S, M.M., and H.O.\ acknowledge funding from the Swedish National
Space Board.
This work has been partially supported by the
Spanish MICINN, within the program CONSOLIDER INGENIO 2010, under grant
``Molecular Astrophysics: The Herschel and Alma Era -- ASTROMOL" (ref.:
CSD2009-00038). R.Sz.\ and M.Sch.\ acknowledge support from grant N 203
393334 from Polish MNiSW. J.C.\ thanks funding from MICINN, 
grant AYA2009-07304. This research was performed, in part, through a 
JPL contract funded by the National Aeronautics and Space Administration.

\end{acknowledgements}

\bibliographystyle{aa}
\bibliography{refs}

\end{document}